\documentclass[twocolumn,showpacs,preprintnumbers,amsmath,amssymb,superscriptaddress,floatfix,nofootinbib]{revtex4}

\usepackage{graphicx}
\usepackage{epsfig}
\usepackage{epstopdf}
\usepackage{hyperref}
\usepackage{amsmath}
\usepackage{amsfonts}
\usepackage{amssymb}

\begin{document}

\title{Revisiting the $\Omega(2012)$ as a hadronic molecule and its strong decays}
\date{\today}
\author{Jun-Xu Lu}
\affiliation{School of Physics $\&$ Beijing Advanced Innovation
Center for Big Data-based Precision Medicine, Beihang University,
Beijing 100191, China}

\author{Chun-Hua Zeng}
\affiliation{Institute of Modern Physics, Chinese Academy of
Sciences, Lanzhou 730000, China} \affiliation{School of Nuclear
Sciences and Technology, University of Chinese Academy of Sciences,
Beijing 101408, China}

\author{En Wang} \email{wangen@zzu.edu.cn}
\affiliation{School of Physics and Microelectronics, Zhengzhou
University, Zhengzhou, Henan 450001, China}

\author{Ju-Jun Xie} \email{xiejujun@impcas.ac.cn}
\affiliation{Institute of Modern Physics, Chinese Academy of
Sciences, Lanzhou 730000, China} \affiliation{School of Nuclear
Sciences and Technology, University of Chinese Academy of Sciences,
Beijing 101408, China} \affiliation{School of Physics and
Microelectronics, Zhengzhou University, Zhengzhou, Henan 450001,
China}

\author{Li-Sheng Geng}\email{lisheng.geng@buaa.edu.cn}
\affiliation{School of Physics $\&$ Beijing Advanced Innovation
Center for Big Data-based Precision Medicine, Beihang University,
Beijing 100191, China} \affiliation{School of Physics and
Microelectronics, Zhengzhou University, Zhengzhou, Henan 450001,
China}

\begin{abstract}

Recently, the Belle collaboration measured the ratios of the
branching fractions of the newly observed $\Omega(2012)$ excited
state. They did not observe significant signals for the
$\Omega(2012) \to \bar{K} \Xi^*(1530) \to \bar{K} \pi \Xi$ decay,
and reported an upper limit for the ratio of the three body decay to
the two body decay mode of $\Omega(2012) \to \bar{K} \Xi$. In this
work, we revisit the newly observed $\Omega(2012)$ from the
molecular perspective where this resonance appears to be a
dynamically generated state with spin-parity $3/2^-$ from the
coupled channels interactions of the $\bar{K} \Xi^*(1530)$ and $\eta
\Omega$ in $s$-wave and $\bar{K} \Xi$ in $d$-wave. With the model
parameters for the $d$-wave interaction, we show that the ratio of
these decay fractions reported recently by the Belle collaboration
can be easily accommodated.

\end{abstract}

\maketitle

\section{Introduction}

In 2018, the Belle collaboration reported an $\Omega^*$ state in the
$\bar{K}\Xi$ invariant mass distributions~\cite{Yelton:2018mag}. The
measured mass and width of the $\Omega^*$ state are $M = 2012.4 \pm
0.7 \pm 0.6$ MeV and $\Gamma = 6.4^{+2.5}_{-2.0} \pm 1.6$ MeV. Such
kind of $\Omega$ excited states have been studied before Belle
collaboration publishes their results. In
Refs.~\cite{Kolomeitsev:2003kt,Sarkar:2004jh,Si-Qi:2016gmh} using
the chiral unitary approach where the coupled channels interactions
of the $\bar{K}\Xi^*(1530)$ and $\eta \Omega$ were taken into
account, the $\Omega$ excited states were investigated. An $\Omega$
excited state with spin-parity $J^P = 3/2^-$ and mass around 2012
MeV can be dynamically generated with a reasonable value of the
subtraction constant~\cite{Si-Qi:2016gmh}. Using a
spin-flavor-$SU(6)$ extended Weinberg-Tomozawa meson-baryon
interaction, the $\Omega$ resonances with $J^P = 1/2^-$, $3/2^-$ and
$5/2^-$ were studied in Ref.~\cite{GarciaRecio:2006bk}. On the other
hand, the $\Omega$ excited states were also investigated in
classical quark
models~\cite{Capstick:1986bm,Loring:2001ky,Pervin:2007wa,Faustov:2015eba}
and in the five-quark
picture~\cite{Yuan:2012zs,An:2013zoa,An:2014lga}, in which, however,
their predicted masses are always much different from the mass
observed by the Belle collaboration. In Ref.~\cite{Chao:1980em},
baryon states with strangeness $-3$ were predicted employing a quark
model with ingredients suggested by QCD, and the mass of one
predicted state with $J^P=3/2^-$ is about $2020$ MeV.

After the observation of the above mentioned $\Omega(2012)$ by the
Belle collaboration~\cite{Yelton:2018mag}, there were many
theoretical studies on its mass, width, quantum numbers and decay
modes. In Refs.~\cite{Aliev:2018syi,Aliev:2018yjo}, the mass and the
two-body strong decays of the $\Omega(2012)$ state were studied by
the QCD sum rule method and it was found that the $\Omega(2012)$ can
be interpreted as a $1P$ orbital excitation of the ground state
$\Omega$ baryon with quantum numbers $J^P = 3/2^-$. In
Refs.~\cite{Xiao:2018pwe,Wang:2018hmi,Liu:2019wdr}, the $\Omega$
excited spectrum and their two body strong decays were evaluated
within a non-relativistic constituent quark potential model, and it
was found that the $\Omega(2012)$ resonance is most likely to be a
$1P$ state with $J^P = 3/2^-$. In Ref.~\cite{Polyakov:2018mow}, the
authors performed a $SU(3)$ flavor analysis of the $\Omega(2012)$
state and discussed its $\bar{K}\Xi^*(1530)$ molecular picture. They
concluded that the preferred quantum numbers of $\Omega(2012)$ are
also $3/2^-$. On the other hand, the mass of the $\Omega(2012)$ is
just a few MeV below the $\bar{K}\Xi^*(1530)$ mass threshold, which
indicates that it could be a possible $\bar{K}\Xi^*(1530)$ molecule
state~\cite{Guo:2017jvc}. Indeed, the hadronic molecule nature of
the $\Omega(2012)$ were investigated in
Refs.~\cite{Valderrama:2018bmv,Lin:2018nqd,Huang:2018wth,Pavao:2018xub},
and these calculations~\footnote{In Ref.~\cite{Huang:2018wth}, the
partial decay width of $\Gamma[\Omega(2012) \to \bar{K} \Xi^*(1530)
\to \bar{K}\pi\Xi] = 3$ MeV was obtained, but this calculation
contained an error. The correct value is 0.8 MeV.} predicted a large
decay width for $\Omega(2012) \to \bar{K} \Xi^*(1530) \to \bar{K}
\pi \Xi$. However, in a very recent measurement of the Belle
collaboration~\cite{Jia:2019eav}, it was found that there is no
significant signals for the $\Omega(2012) \to \bar{K} \Xi^*(1530)
\to \bar{K} \pi \Xi$ decay, and an upper limit was obtained, at the
$90\%$ confidence level, for the ratio of the three body decay to
the two body decay mode of $\Omega(2012) \to \bar{K} \Xi$, $R = {\rm
Br}[\Omega(2012) \to \bar{K}\pi\Xi]/{\rm Br}[\Omega(2012)\to
\bar{K}\Xi]$, which is only $11.9\%$. There are also other
experimental results for the ratios of different final charged decay
modes~\cite{Jia:2019eav}, but because of large background for those
decay channels these values are obtained without including the
constraints of the isospin symmetry~\footnote{Private communications
with Prof. Cheng-Ping Shen.}. Later on, based on the new
measurements by the Belle collaboration~\cite{Jia:2019eav}, the
strong decays of the $\Omega(2012)$ were restudied in
Refs.~\cite{Lin:2019tex,Gutsche:2019imd} within the hadronic
molecular approach. In Ref.~\cite{Lin:2019tex} it concluded that the
$\Omega(2012)$ can be interpreted as the $p$-wave
$\bar{K}\Xi^*(1530)$ molecule state with $J^P = 1/2^+$ or $3/2^+$,
while in Ref.~\cite{Gutsche:2019imd}, it was pointed out that the
$\Omega(2012)$ state contains mixed $\bar{K}\Xi^*(1530)$ and
$\eta\Omega $ hadronic components and the sizable $ \eta\Omega$
hadronic component leads to a suppression of the $\bar{K}\pi \Xi$
decay mode.

The $\Omega(2012)$ state was investigated within a coupled channel
approach in Ref.~\cite{Pavao:2018xub}, in which, in addition to the
interaction of $\bar{K}\Xi^*(1530)$ and $\eta \Omega$ in $s$-wave,
the $\bar{K}\Xi$ in $d$-wave interaction was also taken into
account. The pole position of the $\Omega(2012)$ was well reproduced
in the scattering amplitude. However, the predicted value of $R$ is
about $90\%$~\cite{Pavao:2018xub}, which is much larger than the
experimental measurements~\cite{Jia:2019eav}. Based on the new
measurements of Ref.~\cite{Jia:2019eav}, we follow
Ref.~\cite{Pavao:2018xub} and re-investigate the $\Omega(2012)$
state from the molecular perspective in which the resonance is
dynamically generated from the interactions of $\bar{K}\Xi^*(1530)$,
$\eta \Omega$ and $\bar{K}\Xi$ in coupled channels, with
$\bar{K}\Xi^*(1530)$ and $\eta \Omega$ in $s$-wave and $\bar{K}\Xi$
in $d$-wave. In this work, we determine the unknown parameters
$\alpha$ and $\beta$ introduced in Ref.~\cite{Pavao:2018xub},
fitting to the experimental data, and calculate the partial decay
widths of the two and three body strong decays of $\Omega(2012)$,
with the strong couplings obtained at the pole position of the
state.

The paper is organized as follows. In Section II, we present the
formalism and ingredients of the chiral unitary approach for the
treatment of the $\Omega(2012)$ as a dynamically generated hadronic
state from the interactions of $\bar{K}\Xi^*(1530)$, $\eta \Omega$
and $\bar{K}\Xi$ in coupled channels. Numerical results for the two
and three body strong decays of the $\Omega(2012)$ state and
discussions are given in Section III, followed by a short summary in
the last section.

\section{Formalism and ingredients}

In this section, we briefly review the coupled channel approach to
study the $\Omega(2012)$ state involving the $s$-wave interaction of
$\bar{K}\Xi^*(1530)$, $\eta \Omega$, and $d$-wave interaction of
$\bar{K}\Xi$, although these interactions have been detailed in
Refs.~\cite{Si-Qi:2016gmh,Huang:2018wth,Pavao:2018xub}.

\subsection{Scattering amplitude and the $\Omega(2012)$}

Following Ref.~\cite{Pavao:2018xub}, we denote $\bar{K}\Xi^*(1530)$,
$\eta \Omega$, and $\bar{K}\Xi$ channels by $1$, $2$, and $3$,
respectively, and then the tree level transition amplitudes,
$V_{ij}$ ($i,j = 1, 2, 3$), between each of the two channels are
given by
\begin{eqnarray}
V_{11} &=& V_{22} = V_{33} = 0, \\
V_{12} &=& V_{21} = -\frac{3}{4f_\pi^2}(k^0_1+k^0_2), \\
V_{13} &=& V_{31} = \alpha q_3^2,  \label{Eq:V13} \\
V_{23} &=& V_{32} = \beta q_3^2,   \label{Eq:V23}
\end{eqnarray}
where we take the pion decay constant $f_\pi = 93$ MeV. The $k^0_1$
and $k^0_2$ are the energies of the $\bar{K}$ meson in channel 1 and
$\eta$ meson in channel 2, respectively, which are,
\begin{eqnarray}
k^0_1 &=& \frac{s + m^2_{\bar{K}} - M^2_{\Xi^*}}{2\sqrt{s}}, \\
k^0_2 &=& \frac{s + m^2_{\eta} - M^2_{\Omega}}{2\sqrt{s}}, \\
\end{eqnarray}
with $\sqrt{s}$ the invariant mass of the meson-baryon system.

In addition, $q_3$ is the on-shell momentum of the $\bar{K}$ meson
in channel 3, which reads,
\begin{eqnarray}
q_3 = \frac{\sqrt{[s-(m_{\bar{K}} + M_{\Xi})^2][s - (m_{\bar{K}} -
M_{\Xi})^2]}}{2\sqrt{s}} .
\end{eqnarray}

Then we solve the Bethe-Salpeter equation with the $V_{ij}$ given
above, and obtain the unitarized scattering amplitude $T$:
\begin{equation}\label{BSeq}
T = V + VGT = [1-VG]^{-1}V
\end{equation}
where $G$ is the loop function for each channel and it is a diagonal
matrix containing the meson and baryon propagators. Explicitly
\begin{equation}\label{GLOOP}
  G=\left(
      \begin{array}{ccc}
        G_{11}(\sqrt{s}) & 0 & 0 \\
        0 & G_{22}(\sqrt{s}) & 0 \\
        0 & 0 & G_{33}(\sqrt{s}) \\
      \end{array}
    \right),
\end{equation}
where $G_{ii}(\sqrt{s})$ can be regularized with a cutoff
prescription and the explicit results are~\footnote{More details
about the $d$-wave $\bar{K}\Xi$ loop function can be found in
Ref.~\cite{Pavao:2018xub} and in
Refs.~\cite{Sarkar:2005ap,Roca:2006sz} for the case of the
$\Lambda(1520)$ resonance where the interactions of $\bar{K}N$ and
$\pi\Sigma$ in $d$-wave are included.}:
\begin{equation}\label{GLOOP2}
\begin{split}
  G_{11} = & \int^{\Lambda_1}_0 \frac{d^3q}{(2\pi)^3}\frac{1}{2\omega_1} \frac{M_{\Xi^*}}{E_1}\frac{1}{\sqrt{s} - \omega_1 - E_1 + i\epsilon} \\
  G_{22} = & \int^{\Lambda_2}_0 \frac{d^3q}{(2\pi)^3}\frac{1}{2\omega_2} \frac{M_{\Omega}}{E_2}\frac{1}{\sqrt{s} -\omega_2 - E_2 +i\epsilon} \\
  G_{33} = & \int^{\Lambda_3}_0 \frac{d^3q}{(2\pi)^3}\frac{(q/q_{3})^4}{2\omega_{3}} \frac{M_{\Xi}}{E_{3}}\frac{1}{\sqrt{s} - \omega_{3} -E_{3}+i\epsilon}
\end{split}
\end{equation}
where $E_i$ and $\omega_i$ are the baryon and meson energies for
$i$-th channel. In general, $\Lambda_1$, $\Lambda_2$ and $\Lambda_3$
are different. Yet, to minimize the model parameters, $\Lambda_1 =
\Lambda_2 = 726$ MeV are used in Ref.~\cite{Huang:2018wth} and
$\Lambda_1 = \Lambda_2 = \Lambda_3 = 735$ MeV were used in
Ref.~\cite{Pavao:2018xub}. In this work, we will determine them with
the experimental data of the Belle
collaboration~\cite{Jia:2019eav,Yelton:2018mag}, and discuss them in
the following.

However, since the $\Xi^*(1530)$ resonance has a sizable decay width
and the $\bar{K}\Xi^*(1530)$ mass threshold is close to the mass of
$\Omega(2012)$, the width of $\Xi^*(1530)$ should be considered. For
this purpose, we need to perform a convolution with the spectral
function~\cite{Gamermann:2011mq}
\begin{equation}\label{GLOOP3}
  \overline{G}_{33} = \frac{1}{N}\int_{M_{\Xi^*}-6\Gamma_{\Xi^*}}^{M_{\Xi^*}+6\Gamma_{\Xi^*}} \!\!\! d\tilde{M}\frac{G_{33}(\sqrt{s},\tilde{M}) \tilde{\Gamma}_{\Xi^*}}{(\tilde{M}-M_{\Xi^*})^2 +
  \tilde{\Gamma}^2_{\Xi^*}/4},
\end{equation}
with
\begin{equation}\label{GLOOP4}
N = \int_{M_{\Xi^*}-6\Gamma_{\Xi^*}}^{M_{\Xi^*}+6\Gamma_{\Xi^*}}
d\tilde{M} \frac{\tilde{\Gamma}_{\Xi^*}}{(\tilde{M}-M_{\Xi^*})^2 +
\tilde{\Gamma}^2_{\Xi^*}/4}.
\end{equation}
Note that the range of
$(M_{\Xi^*}-6\Gamma_{\Xi^*},M_{\Xi^*}+6\Gamma_{\Xi^*})$ includes
most of the distribution. Here, the $\tilde{\Gamma}_{\Xi^*}$ is
energy dependent, and its explicit form is given by
\begin{eqnarray}
\tilde{\Gamma}_{\Xi^*}(\tilde{M}) =
\Gamma_{\Xi^*}\frac{M_{\Xi^*}}{\tilde{M}}
 \left ( \frac{\tilde{p}_{\pi}}{p^{\mathrm{on}}_{\pi}} \right )^3,
 \label{Eq:gamrenergy}
\end{eqnarray}
with

\begin{eqnarray}
\tilde{p}_{\pi} \!\! &=& \!\! \frac{\sqrt{[\tilde{M}^2 -(m_{\pi} +
M_{\Xi})^2][\tilde{M}^2 - (m_{\pi} - M_{\Xi})^2]}}{2\tilde{M}} , \nonumber \\
p^{\rm on}_{\pi} \!\! &=& \!\! \frac{\sqrt{[M_{\Xi^*}^2 -(m_{\pi} +
M_{\Xi})^2][M^2_{\Xi^*} - (m_{\pi} - M_{\Xi})^2]}}{2M_{\Xi^*}}.
\nonumber
\end{eqnarray}

In this work, the physical masses and spin-parities of the involved
particles are taken from PDG~\cite{Tanabashi:2018oca}, and tabulated
in Table~\ref{massesparticles}. Note that we take the isospin
averaged values for $m_{K}$, $M_{\Xi^*}$, $M_{\Xi}$ and
$\Gamma_{\Xi^*}$, where we take $\Gamma_{\Xi^*} = 9.5$ MeV.

\begin{table}[htbp]
\centering \caption{Masses and spin-parities of the particles
involved in the present work.}\label{massesparticles}
\begin{tabular}{c|c|c}
\hline\hline
Particle &  Spin-parity ($J^P$) & Mass (MeV) \\
\hline
$\bar{K}$ & $0^-$ & $495.644$ \\
$\eta$ & $0^-$ & $547.862$ \\
$\Xi^*$ & $\frac{3}{2}^+$ & $1533.4$ \\
$\Omega$ & $\frac{3}{2}^+$ & $1672.45$ \\
$\Xi$ & $\frac{1}{2}^+$ & $1318.285$ \\
\hline \hline
\end{tabular}
\end{table}

With this formalism and the former ingredients, one can easily
obtain the scattering matrix $T$. Then one can also look for poles
of the scattering amplitude $T_{ij}$ on the complex plane of
$\sqrt{s}$. The pole, $z_R$, on the second Riemann sheet could be
associated with the $\Omega(2012)$ resonance. The real part of $z_R$
is associated with the mass ($M$) of the state, and the minus
imaginary part of $z_R$ is associated with one half of its width
($\Gamma$). Close to the pole at $z_R = M_R - i \Gamma_R/2$,
$T_{ij}$ can be written as~\footnote{It should be noted that in the
present work the final state interactions of $\bar{K}\Xi^*(1530)$,
$\eta\Omega$, and $\bar{K}\Xi$ are taken into account by solving the
Bethe-Salpeter equation, while they can also be determined more
directly from data (if available), e.g., see,
Refs.~\cite{Au:1986vs,Dai:2014zta}.}
\begin{eqnarray}
T_{ij} = \frac{g_{ii}g_{jj}}{\sqrt{s} - z_R},   \label{tijpole}
\end{eqnarray}
where $g_{kk}$ is the coupling constant of the resonance to the
$k$-th channel . Thus, by determining the residues of the scattering
amplitude $T$ at the pole, one can obtain the couplings of the
resonance to different channels, which are complex in general.

\subsection{The strong decays of $\Omega(2012)$}

\begin{figure}[htbp]
\centering
  \includegraphics[scale=0.65]{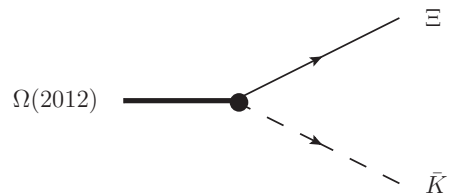}\\
  \caption{The effective $\Omega(2012) \to \bar{K}\Xi$ decay in $d$-wave.}\label{Fig:two-body}
\end{figure}

\begin{figure}[htbp]
\centering
  \includegraphics[scale=0.55]{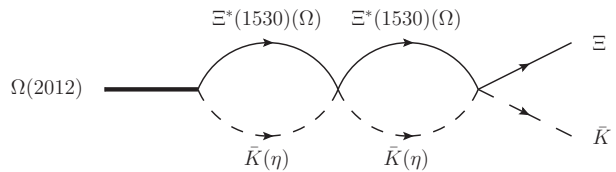}\\
  \caption{The $\Omega(2012) \to \bar{K}\Xi$ decay through the re-scattering of the $\bar{K}\Xi^*(1530)$ and $\eta \Omega$ channels.}\label{Fig:two-body-bubble}
\end{figure}

Since we consider the $s$-wave interactions of the
$\bar{K}\Xi^*(1530)$ and the $\eta \Omega$ channels, the quantum
numbers of the $\Omega(2012)$ should be $J^P =3/2^-$, and it decays
into $\bar{K}\Xi$ in $d$-wave as shown in Fig.~\ref{Fig:two-body},
where the effective interactions are obtained from the $s$-wave
$\Omega(2012)\bar{K}\Xi^*(1530)$ and $\Omega(2012)\eta\Omega$ decays
and the re-scattering of the $\bar{K}\Xi^*(1530)$ and $\eta\Omega$
pairs, which proceed as shown in Fig.~\ref{Fig:two-body-bubble}. In
this respect, the final state interactions of $\bar{K}\Xi \to
\bar{K}\Xi$ is already taken into account in the effective coupling
of the $\Omega(2012)\bar{K}\Xi$ vertex as shown in
Fig.~\ref{Fig:two-body}, since the coupling constant
$g_{\Omega^*\bar{K}\Xi}$~\footnote{In the following, we will use
$g_{\Omega^*\bar{K}\Xi^*} = g_{11}$, $g_{\Omega^*\eta\Omega} =
g_{22}$ and $g_{\Omega^*\bar{K}\Xi} = g_{33}$ for convenience. } is
extracted from the unitarized scattering amplitudes $T$ of
Eq.~\eqref{BSeq}, where the re-scattering of all the three coupled
channels are already included, as shown in
Fig.~\ref{Fig:two-body-bubble}.

Then the partial decay width of the $\Omega(2012) \to \bar{K}\Xi$ is
easily obtained as~\footnote{Note that the $\Omega(2012) \to
\bar{K}\Xi$ decay is in $d$-wave, and it should go on
$q^5_{\bar{K}}$, but the $V_{13}$ and $V_{23}$ potentials of
Eqs.~\eqref{Eq:V13} and \eqref{Eq:V23} incorporate the four extra
powers of $q_{\bar{K}}$.}
\begin{equation}\label{width2b}
\Gamma_{\Omega(2012)\rightarrow \bar{K}\Xi} =
\frac{|g_{\Omega^*\bar{K}\Xi}|^2}{2\pi} \frac{M_{\Xi}}{M}q_{\bar{K}}
,
\end{equation}
where $g_{\Omega^*\bar{K}\Xi}$ is the effective coupling constant of
$\Omega(2012)\bar{K}\Xi$ vertex obtained as explained above, and $M$
is the mass of the obtained $\Omega(2012)$ state, and
\begin{eqnarray}
q_{\bar{K}} = \frac{\sqrt{[M^2 -(m_{\bar{K}} + M_{\Xi})^2][M^2 -
(m_{\bar{K}} - M_{\Xi})^2]}}{2M} .
\end{eqnarray}

\begin{figure}[htbp]
\centering
  \includegraphics[scale=0.6]{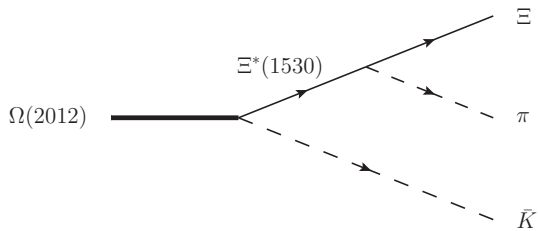}\\
  \caption{Diagram for the three body decay of $\Omega(2012) \to \bar{K}\Xi^*(1530) \to \bar{K}\pi\Xi$.}\label{Fig:three-body}
\end{figure}

For the $\Omega(2012) \to \bar{K}\pi\Xi$ decay, it can proceed via
$\Omega(2012) \to \bar{K}\Xi^*(1530) \to \bar{K}\pi\Xi$. The decay
diagram is shown in Fig.~\ref{Fig:three-body}. And the partial decay
width can be calculated using
\begin{eqnarray}\label{width3b}
 \frac{d\Gamma_{\Omega(2012) \to \bar{K} \pi\Xi}}{dM_{\pi\Xi}}  =  \frac{M_{\pi\Xi}}{\pi^2M} \frac{|g_{\Omega^*\bar{K}\Xi^*}|^2  p_{\bar
  K} \tilde{\Gamma}_{\Xi^*} }{4(M_{\pi\Xi} - M_{\Xi^*})^2 +
  \tilde{\Gamma}_{\Xi^*}^2},
\end{eqnarray}
where $\tilde{\Gamma}_{\Xi^*}$ is dependent on the invariant mass of
$\pi$ and $\Xi$ system, $M_{\pi \Xi}$. And
\begin{eqnarray}
&&p_{\bar{K}} = \frac{\sqrt{[M^2 -(m_{\bar{K}} + M_{\pi\Xi})^2][M^2
- (m_{\bar{K}} - M_{\pi\Xi})^2]}}{2M}. \nonumber
\end{eqnarray}

With all the formulae above, one can easily work out the
$\Gamma_{\Omega(2012)\rightarrow \bar{K}\Xi^{*}  \rightarrow
\bar{K}\pi\Xi}$ performing the integration over $M_{\pi\Xi}$ from
$M_{\Xi}+m_{\pi}$ to $M - m_{\bar K}$.

\section{Numerical results}

To calculate the scattering amplitude $T$, we have to fix the
unknown parameters $\alpha$, $\beta$, and the cutoffs $\Lambda_k$.
Since there are very limited experimental data: the mass and the
width of the $\Omega(2012)$ and the upper limit of the ratio $R$, we
will take the same value for $\Lambda_1 =\Lambda_2 = \Lambda_3 =
q_{\rm max}$. Even so, we still have three free parameters, and
there are only two experimental data plus one more constraint, the
upper limit of the ratio $R < 11.9\%$.

Varying the unknown model parameters of $\alpha$, $\beta$ and
$q_{\rm max}$, we find that one can reproduce the mass and width of
$\Omega(2012)$ and the upper limit $R < 11.9\%$ with the following
range of the model parameters~\footnote{In fact, we find that one
can only determine the relative sign between $\alpha$ and $\beta$,
rather than their absolute signs. In this work, we take negative
sign for $\alpha$ and positive sign for $\beta$.}:
\begin{eqnarray}
&&  \alpha < -5 \times 10^{-8}\mathrm{MeV}^{-3},~\beta>15 \times 10^{-8}\mathrm{MeV}^{-3}, \label{rangepara} \\
&&  q_{\rm max} >
  720~\mathrm{MeV}.
\end{eqnarray}

To minimize the number of the free parameters, we fix $q_{\rm max} =
735$ (Set I), $750$ (Set II), $800$ (Set III), $850$ (Set IV), and
$900$ MeV (Set V), and determine $\alpha$ and $\beta$ by fitting
them to the experimental data. Since we only know the upper limit of
$R$, it is difficult to perform a $\chi^2$ fit to it. Technically,
one can define
\begin{eqnarray}
\chi^2 = \left (\frac{M^{\rm th} - M^{\rm exp}}{\Delta M^{\rm exp}}
\right )^2 + \left (\frac{\Gamma^{\rm th} - \Gamma^{\rm exp}}{\Delta
\Gamma^{\rm exp}} \right )^2,
\end{eqnarray}
where $M^{\rm th}$ and $\Gamma^{\rm th}$ are evaluated at the pole
position of $T$, and we take $M^{\rm exp} = 2012.4$ MeV, $\Delta
M^{\rm exp} = 0.9$ MeV, $\Gamma^{\rm exp} = 6.4$ MeV, and
$\Delta\Gamma^{\rm exp} = 3.0$ MeV as measured by the Belle
collaboration~\cite{Yelton:2018mag}. Then we vary firstly the values
of $\alpha$ and $\beta$ in the range as in Eq.~\eqref{rangepara}. If
the obtained mass and width of $\Omega(2012)$, and $R$ are in
agreement with the experimental values within errors, we call that a
best fit. In this way, we obtain sets of the fitted parameters
($\alpha$, $\beta$) with different best $\chi^2_{\rm best}$. The
fitted parameters corresponding to the minimum $\chi^2_{\rm min}$
that we get from the best fits are: $(\alpha, \beta) = (-8.0,
17.6)$, $(-11.1, 19.5)$, $(-18.5, 21.2)$, $(-21.8, 20.6)$, and
$(-22.0, 18.2) \times 10^{-8}{\rm MeV}^{-3}$ for sets I, II, III, IV
and V, respectively, which are listed in
Table~\ref{fittedparameters}. We will take these values as the
central values of parameters $\alpha$ and $\beta$. In addition, with
all the fitted parameters of $\alpha$ and $\beta$ with the
$\chi^2_{\rm best}$ fit, we search for the minimal values of the
ratio $R$, which are $9\%$, $8\%$, $7\%$, $5\%$ and $4\%$ for sets
I, II, III, IV and V, respectively. It is worth to mention that, in
Ref.~\cite{Gutsche:2019imd}, the minimal value of $R$ could be zero.

Next, we collect these sets of the fitted parameters, such that the
corresponding $\chi^2_{\rm best}$ are below $\chi^2_{\rm min} + 1$.
With these collected best fitted parameters, we obtain the standard
deviations of parameters $\alpha$ and $\beta$, which are quoted in
Table~\ref{fittedparameters} as their errors. In the same table, we
also show the obtained pole positions of the $\Omega(2012)$ state
and the couplings with the best fitted parameters.~\footnote{The
theoretical uncertainties for $M_R$, $\Gamma_R$ and these couplings
constants shown in Table~\ref{fittedparameters} are obtained in the
same way.} Note that the pole position of $\Omega(2012)$ is about
$17$ MeV below the mass threshold of $\bar{K}\Xi^*(1530)$, and is
far from the thresholds of $\eta \Omega$ and $\bar{K}\Xi$. According
to the "criteria" for the innner structure of a pole as proposed in
Refs.~\cite{Morgan:1992ge,Dai:2011bs,Dai:2012kf} for the scalar
mesons and in Ref.~\cite{Kuang:2020bnk} for the pentaquark states,
we conclude that the $\Omega(2012)$ is a molecular hadronic state,
dynamically generated from the interactions of $\bar{K}\Xi^*(1530)$
and $\eta \Omega$ in $s$-wave and $\bar{K}\Xi$ in $d$-wave.

\begin{table*}[htbp]
\centering \caption{Determined values of the unknown parameters in
this work. We also give the pole positions $(M_R,\Gamma_R)$ of the
$\Omega(2012)$ and the couplings to different channel obtained with
the central values of these fitted
parameters.}\label{fittedparameters}
\begin{tabular}{c|c|c|c|c|c|c|c}
\hline\hline Set & $q_{\rm max}$ (MeV) & $\alpha$ ($10^{-8}$ ${\rm MeV}^{-3}$) & $\beta$ ($10^{-8}$ ${\rm MeV}^{-3}$) & $(M_R, \Gamma_R)$ (MeV) & $|g_{\Omega^*\bar{K}\Xi^*}|$  & $|g_{\Omega^*\eta\Omega}|$ & $|g_{\Omega^*\bar{K}\Xi}|$   \\
\hline
I   & $735$   & $-6.6 \pm 0.8$   & $16.5 \pm 0.8$  & $(2012.3 \pm 0.4, 8.3 \pm 0.6)$  & $1.83 \pm 0.02$  & $3.35 \pm 0.06$ & $0.42 \pm 0.02$ \\
II  & $750$   & $-9.9 \pm 0.5$   & $18.5 \pm 0.5$  & $(2012.2 \pm 0.4, 7.8 \pm 0.8)$  & $1.80 \pm 0.01$  & $3.46 \pm 0.06$ & $0.41 \pm 0.03$ \\
III & $800$   & $-17.5 \pm 0.6$  & $20.6 \pm 0.5$  & $(2012.4 \pm 0.5, 6.4 \pm 1.3)$  & $1.58 \pm 0.02$  & $3.60 \pm 0.04$ & $0.37 \pm 0.04$ \\
IV  & $850$   & $-20.2 \pm 1.0$  & $19.6 \pm 0.8$  & $(2012.4 \pm 0.5, 6.4 \pm 1.1)$  & $1.39 \pm 0.03$  & $3.78 \pm 0.04$ & $0.37 \pm 0.03$ \\
V   & $900$   & $-20.8 \pm 1.7$  & $17.5 \pm 1.1$  & $(2012.4 \pm 0.5, 6.4 \pm 1.3)$  & $1.25 \pm 0.04$  & $3.85 \pm 0.04$ & $0.37 \pm 0.04$ \\
\hline \hline
\end{tabular}
\end{table*}

In addition, with the mass of the $\Omega(2012)$ state and coupling
constants obtained from the best fit, we calculate the partial decay
widths of $\Omega(2012) \to \bar{K}\pi\Xi$ and $\Omega(2012) \to
\bar{K}\Xi$, and also their ratio $R$.~\footnote{The $R$ is obtained
with the central values of ${\rm Br}[\Omega(2012) \to
\bar{K}\pi\Xi]$ and ${\rm Br}[\Omega(2012) \to \bar{K}\Xi]$ shown in
Table~\ref{predictions}.} We show these results in
Table~\ref{predictions}. Note that to get the uncertainties of ${\rm
Br}[\Omega(2012) \to \bar{K}\pi\Xi]$ and ${\rm Br}[\Omega(2012) \to
\bar{K}\Xi]$, we have considered that the sum of them should be less
than one. From these results, one can easily find that the sum of
the branching fractions of $\Omega(2012) \to \bar{K}\pi\Xi$ and
$\Omega(2012) \to \bar{K}\Xi$ is more than $95\%$, which indicates
that the other decay modes and other strong decay mechanisms of
$\Omega(2012)$ are small, such as those of the triangle mechanisms
of Refs.~\cite{Huang:2018wth,Lin:2019tex}.

\begin{table*}[htbp]
\centering \caption{The predicted results for the two and three body
strong decays of the $\Omega(2012)$ with the fitted parameters given
in Table~\ref{fittedparameters}.} \label{predictions}
\begin{tabular}{c|c|c|c|c|c}
\hline\hline Set & $\Gamma_{\Omega(2012) \to \bar{K}\pi\Xi}$  (MeV)
& $\Gamma_{\Omega(2012) \to \bar{K}\Xi}$  (MeV) & ${\rm
Br}[\Omega(2012) \to \bar{K}\pi\Xi]$ & ${\rm Br}[\Omega(2012) \to
\bar{K}\Xi]$ & $R$ \\
\hline
I      & $0.87 \pm 0.03$   & $7.32 \pm 0.64$ & $(10.5^{+0.5}_{-0.8})\%$ & $(88.4 ^{+0.5}_{-1.5})\%$  & $11.88\%$  \\
II     & $0.84 \pm 0.04$   & $6.96 \pm 0.63$ & $(9.5^{+0}_{-1.0})\%$ & $(90.5^{+0}_{-2.6})\%$  & $10.50\%$  \\
III    & $0.66 \pm 0.02$   & $5.57 \pm 1.37$ & $(10.3^{+1.6}_{-1.7})\%$  & $(86.5^{+1.6}_{-2.9})\%$  & $11.90\%$  \\
IV     & $0.51 \pm 0.03$   & $5.66 \pm 1.07$ & $(7.9^{+1.9}_{-1.5})\%$  & $(88.2^{+1.9}_{-1.6})\%$  & $9.00\%$   \\
V      & $0.41 \pm 0.03$   & $5.73 \pm 1.25$ & $(6.5^{+1.7}_{-1.9})\%$  & $(90.0^{+1.7}_{-2.2})\%$  & $7.22\%$   \\
\hline \hline
\end{tabular}
\end{table*}

\begin{figure}[htbp]
\centering \hspace{1.cm}
  \includegraphics[scale=0.4]{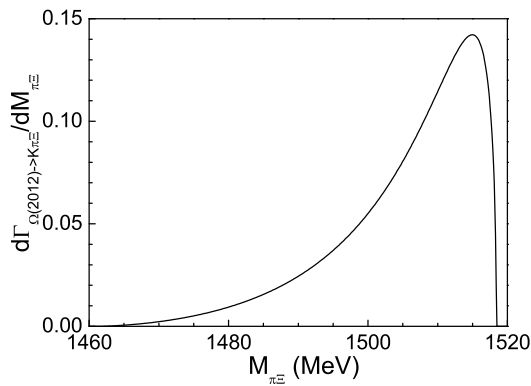} \\
  \caption{The $\pi\Xi$ invariant mass distribution of the three body decay of $\Omega(2012) \to \bar{K}\Xi^*(1530) \to \bar{K}\pi\Xi$. The numerical results are obtained with the parameters of set I.} \label{Fig:dgdm}
\end{figure}

Finally, we pay attention to the $\pi\Xi$ invariant mass
distributions of the $\Omega(2012) \to \bar{K}\Xi^* \to
\bar{K}\pi\Xi$ decay. The theoretical calculations with the
parameters of Set I are shown in Fig.~\ref{Fig:dgdm}. One can see
that, because of the phase space limitations,
$d\Gamma_{\Omega(2012)\to \bar{K}\pi\Xi}/dM_{\pi\Xi}$ peaks around
$M_{\pi\Xi} = 1515$ MeV, which is lower than the mass of
$\Xi^*(1530)$. Since the width of $\Xi^*(1530)$ is narrow, it is
easy to see that the $\Omega(2012) \to \bar{K}\Xi^*(1530) \to
\bar{K}\pi\Xi$ decay will be much suppressed due to the highly
off-shell effect of the $\Xi^*(1530)$ propagator, though the
coupling of $\Omega(2012)$ to the $\bar{K}\Xi^*(1530)$ channel is
strong.

\section{Summary}

Based on the recent measurements by the Belle
collaboration~\cite{Jia:2019eav}, where they did not observe
significant signals for the $\Omega(2012) \to \bar{K} \Xi^*(1530)
\to \bar{K} \pi \Xi$ decay, we revisit the $\Omega(2012)$ state from
the molecular perspective in which this resonance appears to be
dynamically generated from the coupled channels interactions of the
$\bar{K} \Xi^*(1530)$ and $\eta \Omega$ in $s$-wave and $\bar{K}
\Xi$ in $d$-wave. In such a scenario, the $\Omega(2012)$ is
interpreted as a $3/2^-$ molecule state. We studied the two and
three body strong decays of $\Omega(2012)$, within the model
parameters for the $d$-wave interaction, it is shown that the
experimental properties of the $\Omega(2012)$ reported recently by
the Belle collaboration can be easily accommodated. More and precise
experimental measurements on the strong decays of the $\Omega(2012)$
would be very useful to better understand its nature.

\section*{Acknowledgments}

XJJ and LSG would like to thank Prof. Eulogio Oset and Prof.
Cheng-Ping Shen for fruitful discussions. This work is partly
supported by the National Natural Science Foundation of China under
Grant Nos. 11735003, 11975041, 11961141004, 11565007, 11847317,
11975083, 1191101015 and the Youth Innovation Promotion Association
CAS (2016367). It is also supported by the Key Research Projects of
Henan Higher Education Institutions under No. 20A140027, the
Academic Improvement Project of Zhengzhou University.

\end{document}